\documentclass[aps,superscriptaddress,twocolumn,pre,longbibliography,floatfix]{revtex4-1}
\usepackage{amsmath,amsthm,amsfonts,amssymb,bm,graphicx,color,mathpazo,times, mathtools, dsfont}
\usepackage[colorlinks={true}, citecolor={blue}, filecolor={blue}, linkcolor={blue}, urlcolor={blue}]{hyperref}
\usepackage[caption=false]{subfig}
\usepackage{soul}
\usepackage{braket}
\usepackage[toc,page,titletoc]{appendix}
\usepackage[capitalise,nameinlink]{cleveref}
\newcommand{\abs}[1]{\left\vert#1\right\vert}

\crefname{supp}{Supplement}{Supplements}
\allowdisplaybreaks
\begin{document}
\preprint{APS/123-QED}
\title{Minimal quantum heat manager boosted by bath spectral filtering}
\author{M. Tahir Naseem}
\affiliation{Department of Physics, Ko\c{c} University, 34450 Sariyer, Istanbul Turkey}

\author{Avijit Misra}
\affiliation{Department of Chemical and Biological Physics, Weizmann Institute of Science, 
Rehovot 7610001, Israel}

\author{\"{O}zg\"{u}r E. M\"{u}stecapl{\i}o\u{g}lu}
\email{omustecap@ku.edu.tr}
\affiliation{Department of Physics, Ko\c{c} University, 34450 Sariyer, Istanbul Turkey}

\author{Gershon Kurizki}
\affiliation{Department of Chemical and Biological Physics, Weizmann Institute of Science, 
Rehovot 7610001, Israel}
\date{\today}
\begin{abstract}
We reveal  the potentially important role of a general mechanism in quantum heat management schemes, namely, spectral filtering of the coupling between the heat baths in the setup and the quantum system that controls the heat flow. Such filtering is enabled by interfaces between the system and the baths by means of harmonic-oscillator modes whose resonant frequencies and coupling strengths are used as control parameters of the system-bath coupling spectra. We show that this uniquely quantum-electrodynamic mechanism, here dubbed bath spectral filtering, boosts the performance of a minimal quantum heat manager comprised of two interacting qubits or an analogous optomechanical system, allowing this device to attain either perfect heat diode action or  strongly enhanced heat transistor action.
\end{abstract}   
\maketitle
\section{\label{sec:Intro}Introduction}
There has been tremendous upsurge in theoretical activity related to heat management in quantum systems, particularly heat flow rectification or amplification~\cite{PhysRevLett.94.034301, PhysRevLett.99.027203, PhysRevLett.100.105901, PhysRevLett.107.173902, PhysRevE.89.062109, Martínez-Pérez2015, PhysRevE.94.042135, PhysRevE.95.022128, Karimi_2017, Ronzani2018,PhysRevE.99.042121,PhysRevLett.116.200601, Zhang_2018, PhysRevA.97.052112, PhysRevE.99.032114, PhysRevE.99.042102, PhysRevE.99.062123,Motz_2018,senior2019heat}. This activity has been mainly motivated by interest in  potential quantum technological applications, but also by the quest for new insights into quantum thermodynamics. However, the lack of fundamental, general principles of quantum heat management is underscored by the abundance of models and diverse approaches to  the subject. Such principles are needed not only for deeper conceptual understanding of quantum thermodynamics, but also as guidance for the design of a quantum device capable of near-perfect execution of the aforementioned functionalities, namely,  heat-flow rectification, known as heat-diode (HD) action~\cite{PhysRevE.89.062109, PhysRevE.94.042135, PhysRevE.95.022128,PhysRevLett.107.173902, Karimi_2017, Ronzani2018,PhysRevE.99.042121,Motz_2018,senior2019heat}, as well as heat-flow amplification with negative differential heat resistance, alias heat transistor (HT) action~\cite{PhysRevLett.116.200601, Zhang_2018, PhysRevA.97.052112, PhysRevE.99.032114, PhysRevE.99.042102, PhysRevE.99.062123}.  
 
 Here we reveal  the potentially important role of a general mechanism that has hitherto been little invoked 
 \cite{Motz_2018,senior2019heat} in quantum heat management schemes, namely, spectral filtering of the coupling between the heat baths in the setup and the quantum system that controls the heat flow. Such filtering is enabled by interfaces between the system and the baths by means of harmonic-oscillator modes whose resonant frequencies and coupling strengths are used as control parameters of the system-bath coupling spectra. 
  We show that this mechanism (Sec.~\ref{sec:model}), here dubbed bath spectral filtering (BSF), boosts the performance of a minimal quantum heat manager comprised of two interacting qubits or an analogous optomechanical system, allowing this device to attain either perfect HD action (Sec.~\ref{sec:Rect}) or enhanced HT action (Sec.~\ref{sec:Ampli}).  We stress that the BSF is a genuinely quantum electrodynamic effect, which stands in contrast to most mechanisms employed  in existing  quantum heat management schemes that have classical counterparts. Feasible experimental setups are proposed (Sec.~\ref{sec:ExpReal}) and the relevant derivations are outlined (App.~\ref{AppendixA},\ref{AppendixB},\ref{AppendixC},\ref{AppendixD})
%
\begin{figure}[t]
\centering
\includegraphics[width=0.350\textwidth]{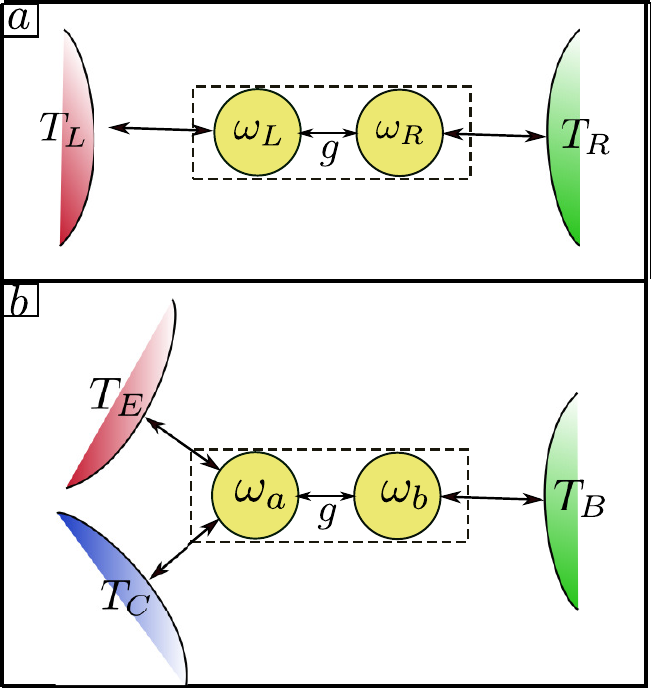}
\caption{\label{fig:BareQubits}(Color online) Schematic diagram of (a) heat diode (HD), (b) heat transistor (HT) based on two coupled two-level systems (TLS) or two harmonic oscillators with an anisotropic interaction strength $g$.  For the HD, the frequency of the subsystem `L' (`R') is $\omega_{L}$ ($\omega_{R}$), and it is coupled to a thermal bath at temperature $T_{L}$ ($T_{R}$). For the HT,  the sub-system `a' is coupled to two thermal baths at temperatures $T_{E}$ and $T_{C}$, while the sub-system `b' is coupled to a single thermal bath at temperature $T_{B}$. These baths are independent and may have any distinct non-negative temperatures.}
\end{figure}
\section{Heat management with BSF}\label{sec:model}

Consider a  multilevel system $S$ that is sandwiched between heat baths on its left (L) and right (R) sides. If these baths differ only in temperature, can such a system control or manage the heat flow between the baths? The expressions for the heat currents through the system are given in the Markovian approximation by \cite{kosloff2014quantum, Implementations,Gelbwaser_2015}
\begin{equation}\label{Eq:HeatCurrent1}
\mathcal{J}_{\alpha} = \text{Tr}\{(\mathcal{L}_{\alpha}\rho)H_{S}\}.
\end{equation}
Here $\mathcal{L}_{\alpha}$ is the Lindbladian corresponding to the bath $\alpha$,  where $\alpha=L, R,$  and the time evolution of $\rho$ is given by
\begin{equation}
\dot{\rho} = \mathcal{L}_{L}(\rho) +\mathcal{L}_{R}(\rho).
\end{equation}
The Lindbladians have the form
\begin{equation}\label{Eq:lind}
\mathcal{L}_{\alpha}= \sum_i [ G_{\alpha}(\omega_i)\hat{\mathcal{D}}[A_{\omega_i}]
+ G_{\alpha}(-\omega_i)\hat{\mathcal{D}}[A_{\omega_i}^\dagger]],
\end{equation}
where for any pair of noncommuting operators $O$ and $O^\dagger$ the dissipator is given by
\begin{equation}\label{dissipator}
\hat{\mathcal{D}}[{O}] = \frac{1}{2}\left(2{O}\rho O^{\dagger} - {O}^{\dagger}{O}{\rho} - {\rho}{O}^{\dagger}{O}\right),
\end{equation}
and $A_{\omega_i}$ and $A_{\omega_i}^\dagger$ are the lowering and raising operators corresponding to the eigenstates  of the system with energy difference $\omega_i$.
The system -bath coupling spectra are given by~\cite{Gelbwaser_2015, PhysRevE.90.022102, Gelbwaser-Klimovsky2015, PhysRevE.87.012140}
\begin{eqnarray}\label{eq:SRF}
G_{\alpha}(\omega)=
\begin{cases}
\kappa_{\alpha}(\omega)[1 + \bar{n}_{\alpha}(\omega)] &\omega> 0, \\
\kappa_{\alpha}(\abs{\omega})\bar{n}_{\alpha}(\abs{\omega}) &\omega< 0, \\
0 &\omega = 0.
\end{cases}
\end{eqnarray}

Here,  $\bar{n}_\alpha (\omega)= 1/\big[\exp{(\omega/T_{\alpha})} - 1\big]$  is the mean excitation  number (number of quanta) in the thermal baths and $\kappa_\alpha (\omega)$ is the coupling strength of the system to the respective bath. 

Let us assume that the two heat baths have the same characteristics, so that $\mathcal{L}_L$
and $\mathcal{L}_R$ differ only in temperature. The minimal model for $S$ suitable for heat management is then~\cite{PhysRevA.83.032105}
\begin{equation}\label{eq:H1}
\hat{H}_{1} = \frac{\omega_{L}}{2}\hat{\sigma}_{L}^{z} + \frac{\omega_{R}}{2}\hat{\sigma}_{R}^{z} + g\hat{\sigma}_{L}^{z}\hat{\sigma}_{R}^{x},
\end{equation}
 where z and x label the corresponding Pauli matrices. Upon diagonalizing this Hamiltonian, we obtain
\begin{equation}\label{eq:diaghamil}
{\tilde{H}}_{1} = \frac{\omega_{L}}{2}{\tilde{\sigma}}_{L}^{z} + \frac{\Omega}{2}{\tilde{\sigma}}_{R}^{z},
\end{equation}
where $\Omega = \sqrt{\omega_{R}^{2} + 4g^{2}}$. The transformed Pauli matrices are denoted by ${\tilde{\sigma}}_\alpha^\beta$ where $\beta=x,y,z$. The subsystems L and R are coupled to corresponding baths.
In this transformed basis, the Lindblad superoperator for the right-hand bath reads (App.~\ref{AppendixA})
\begin{eqnarray}
\label{eq:L_L}
\hat{ \mathcal{L}}_{R}&=& G_{R}(\Omega)\cos^{2}\theta\hat{\mathcal{D}}[{\tilde{\sigma}}_{R}^{-}]
+ G_{R}(-\Omega)\cos^{2}\theta\hat{\mathcal{D}}[{\tilde{\sigma}}_{R}^{+}], \label{eq:L_R}
\end{eqnarray}
with $\theta = \arctan(2g/\omega_{R})$. Thanks to $\theta$ being nonzero, the Lindblad superoperators are not L-R interchangeable even if the temperatures are interchanged.

Another minimal model for heat management in this scenario is the optomechanical system (OMS) associated with photon-phonon interactions \cite{Kurizki_2015,Gelbwaser-Klimovsky2015}     
\begin{equation}\label{eq:H3}
\hat{H}_{2} = \omega_{L}\hat{a}^{\dagger}\hat{a}+\omega_{R}\hat{b}^{\dagger}\hat{b}-g\hat{a}^{\dagger}\hat{a}(\hat{b}+\hat{b}^{\dagger}).
\end{equation}
In the limit of small excitation numbers this model becomes isomorphic to the  two-qubit model, so that it suffices to consider the latter in what follows. 

The tacit assumption behind all treatments of quantum heat management thus far has been that the couplings $\kappa_\alpha (\omega)$ are spectrally flat, an assumption dubbed flat spectral density (FSD).  
We stress here that FSD can put severe constraints on heat control  by coupled subsystems, as in the minimal models above. This is explained in detail in Secs.~\ref{sec:Rect} and~\ref{sec:Ampli}. Can one overcome these restrictions and attain better heat control for weakly-asymmetric L-R subsystems? 

The remedy proposed here is bath-spectral filtering (BSF). Our central point is that the ability to controllably shape $\kappa_\alpha (\omega)$, thus abandoning the FSD constraint, provides a key resource for 
heat management. This ability comes about if each bath is supplemented with a harmonic-oscillator (HO-) mode that serves as an interface between the bath and the respective subsystem, e.g. qubit. As first shown in~\cite{JMO94} and subsequently employed in the analysis of quantum heat machines~\cite{PhysRevE.90.022102, Gelbwaser-Klimovsky2015, PhysRevE.87.012140,Ghosh2018}, a filter HO mode with resonant frequency $\tilde{\omega} _{ \alpha}$ that is coupled to the qubit with strength $\eta _{ \alpha}$ and to the bath via coupling spectrum ${G}_\alpha (\omega)$, yields the following modified (filtered) qubit-bath coupling spectrum
\begin{equation}\label{eq:BathFilter}
\tilde{G}_\alpha = \frac{\eta_{\alpha}}{\pi}\frac{(\pi G_{\alpha}(\omega))}{[\omega-\big(\tilde{\omega}_{\alpha}+\Delta_{\alpha}(\omega)\big)]^2+(\pi G_{\alpha}(\omega))^2},
\end{equation}
where $G_{\alpha}$ is the unfiltered coupling spectrum, and
\begin{equation}
\Delta_{\alpha}(\omega) = P \bigg[\int^{\infty}_{o} d\omega^{'} \frac{G_{\alpha}(\omega^{'})}{\omega - \omega^{'}}\bigg],
\end{equation}
$P$,
$\Delta_\alpha (\omega)$ being, respectively,  the principal value and the bath-induced Lamb shift. Similar results are  shown in App.~\ref{AppendixB} to hold  for OMS. The filtered spectrum (Eq.~\ref{eq:BathFilter}) can be drastically different from its original (unfiltered)  counterpart. In general, the spectral shape of the filtered spectrum is a {\it skewed-Lorentzian}~\cite{JMO94}. In the case of unfiltered FSD, the filtered counterpart is a regular Lorentzian \cite{Motz_2018,senior2019heat}
\begin{eqnarray}\label{eq:W_01}
\tilde{G}_\alpha=\frac{\eta_{\alpha}}{\pi}\frac{\kappa^{2}_{\alpha}}{(\omega-\omega_\alpha)^2+(\pi\kappa_\alpha)^2},
\end{eqnarray}
whose width and center are, respectively, the controllable $\kappa _{\alpha}$ and $\omega _{\alpha}$. However, Eq.~(\ref{eq:W_01}) may not suffice for our purposes, since the tails of a regular Lorentzian fall off too slowly with frequency. Instead, we require BSF that yields a strongly asymmetric skewed-Lorentzian with fast spectral drop-off on one wing.

 In the case of coupled qubits, one filtered bath couples to the transition at frequency  $\Omega$, while the other to the transition at frequencies $\omega_L$ and $\omega_L+\Omega$, 
provided the coupling spectra are filtered to be skewed-Lorentzians that satisfy
\begin{eqnarray}\label{eq:Rspectrum}
G_L (\omega_L +\Omega)&\gg&  G_R (\omega_L +\Omega),\\\nonumber
G_L (\omega_L)&\gg&  G_R (\omega_L),\\\nonumber
G_L (\Omega)&\ll& G_ R (\Omega).
\end{eqnarray}
Only such BSF here can separate the coupling spectra of the L-and R-subsystems to their respective baths.\\  
\begin{figure}[t]
\centering
\includegraphics[width=0.40\textwidth]{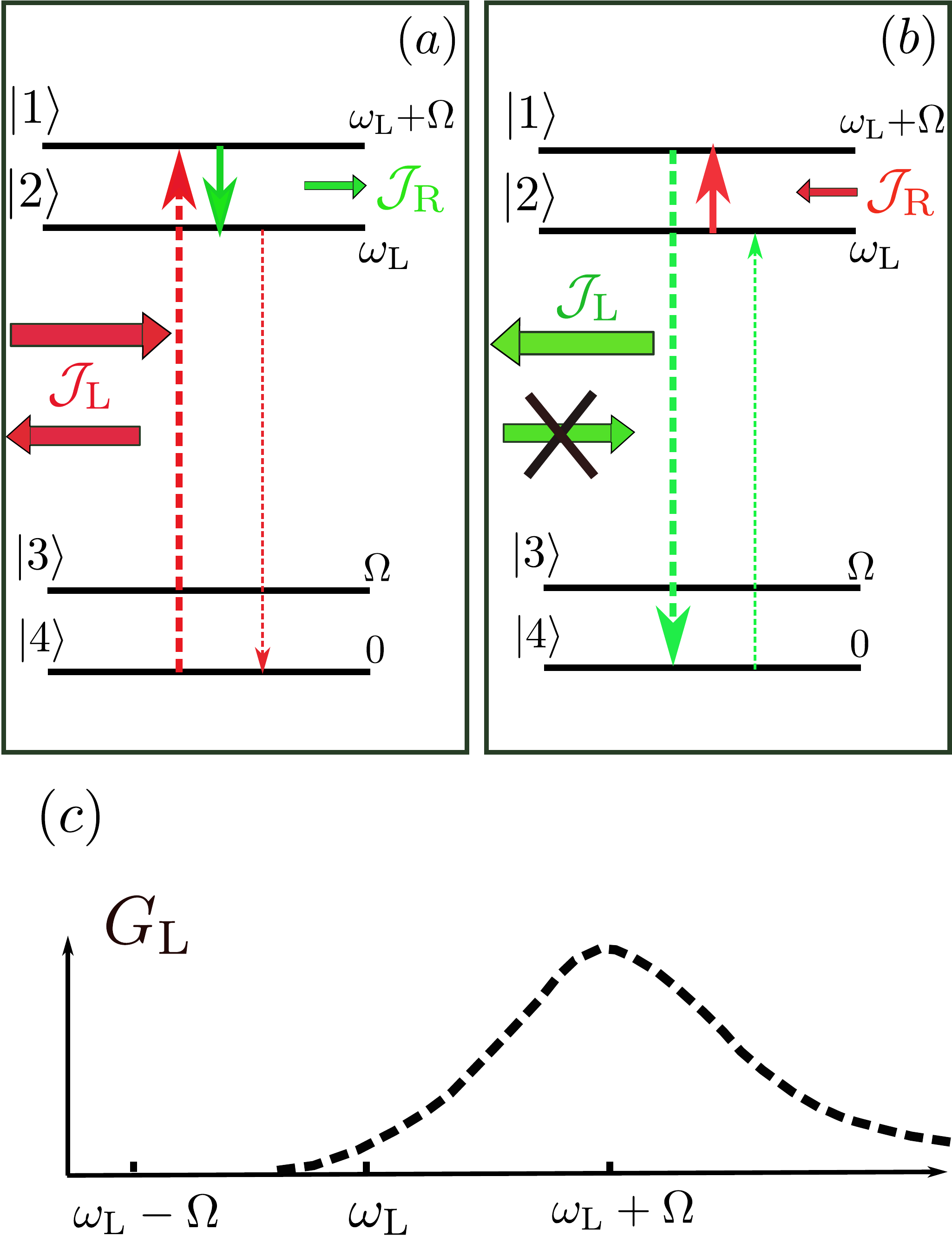}
\caption{(Color online) The process that rectifies heat current in an HD based on two coupled TLS. In panel (a) the left bath is hotter than the right bath, and in panel (b) it is the opposite. The dashed and dotted lines correspond to transitions induced by the right and left baths, respectively. Solid arrow thickness represents the transition rate. The transition at $\omega_{L}$ is weak due to the choice of a filtered left bath spectrum, whereas the right bath has FSD.
Heat can flow from left to right via the only possible Raman cycle (4124) in (a), whereas the opposite cycle (4214) in (b) is inhibited because the cold bath cannot excite the $|4\rangle \to |2\rangle $ transition. Accordingly, there is no heat flow in panel (b), and our HD gives perfect rectification. (c) Filtered spectral response function of the L bath with $G(\omega_{L}-\Omega)=0$ is a strongly skewed-Lorentzian obtained according to Eqs.~(\ref{eq:BathFilter}), ~(\ref{eq:Rspectrum}).}
\label{fig:qqRectMechanism}
\end{figure}
\begin{figure}[t]
\centering
\hspace*{-0.60cm}
\includegraphics[width=0.5\textwidth]{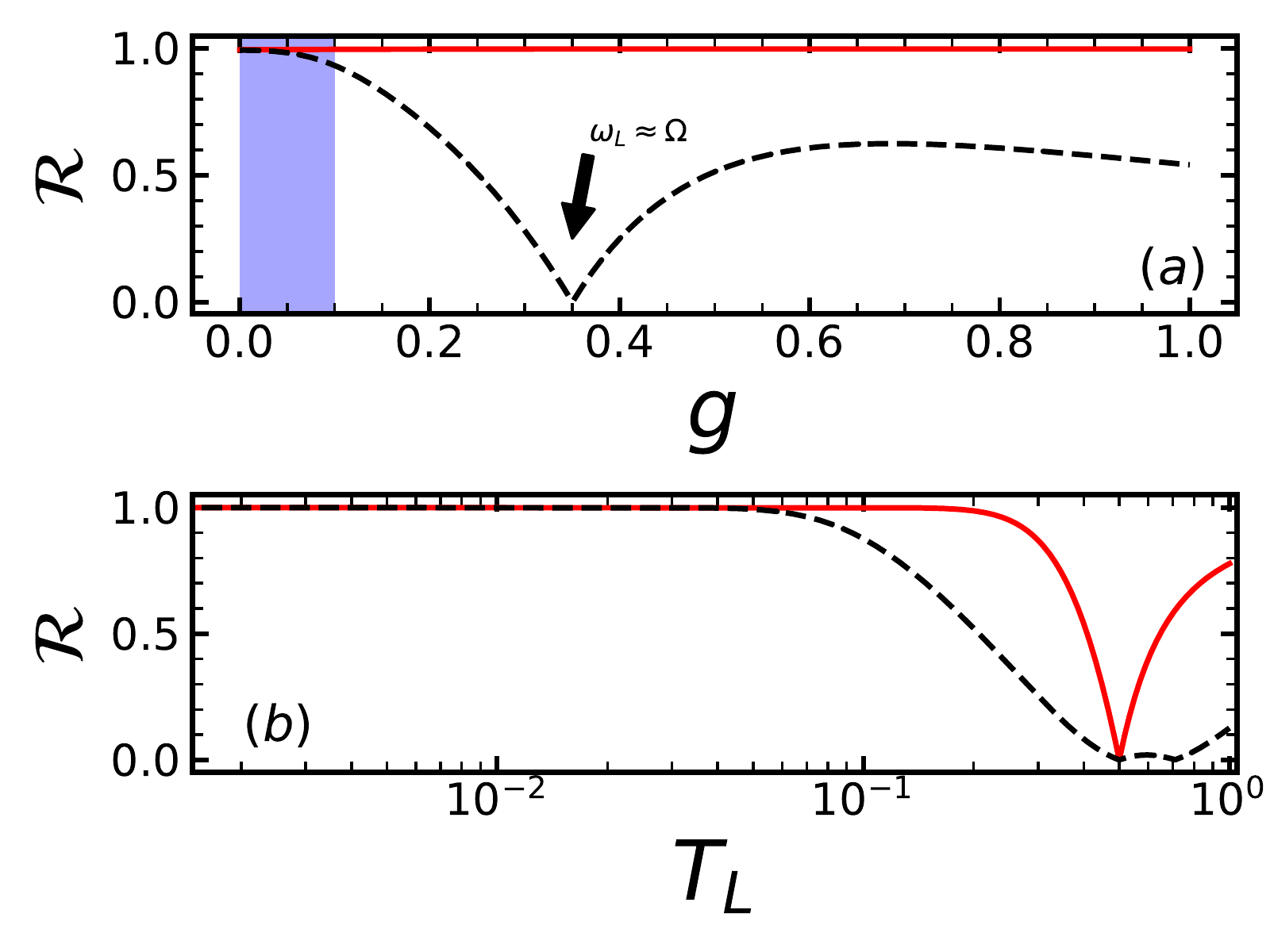}
\caption{(Color online)~{\it
{Rectification}} $\mathcal{R}$ as a function of the (a) coupling strength $g$, (b) temperature $T_{L}$ for the coupled TLS. The solid and dashed lines are for the filtered and unfiltered L bath spectrum, respectively. 
The L bath has filtered spectral density as shown in Fig.~\ref{fig:qqRectMechanism}(c), and the right bath has FSD. The shaded region is for $\omega_{L}\gg\Omega$.
Parameters: $\omega_{L}=1$, $\omega_{R}=0.1$, $\kappa_{L}=\kappa_{R}=0.001$, (a) $T_{L}=2$, $T_{R}=0.2$, and (b) $g=0.35$, $T_{R}=0.5$. All the system parameters are scaled with $\omega_{L}/2\pi=10$ GHz.}
\label{fig:qRect}
\end{figure}
\section{Rectification with BSF}\label{sec:Rect}
Heat flow rectification between two baths~\cite{PhysRevE.89.062109, PhysRevE.94.042135, PhysRevE.95.022128,PhysRevLett.107.173902, Karimi_2017, Ronzani2018,PhysRevE.99.042121} is quantified by the rectification factor
\begin{equation}\label{Eq:Rect}
\mathcal{R} = \frac{\abs{\mathcal{J}_{R}(T_{R},T_{L}) + \mathcal{J}_{R}(T_{L},T_{R})}}{Max(\abs{\mathcal{J}_{R}(T_{R},T_{L})}, \abs{\mathcal{J}_{R}(T_{L},T_{R}})},
\end{equation}
where $\mathcal{J}_{L}$ ($T_R, T_L$) is the heat current for $T_L > T_R$ ß. The rectification factor varies between 1 for perfect rectification and 0 for complete reciprocity of the heat flows. It is clear from Eq.~(\ref{Eq:lind}) that if $\mathcal{L}_L$
and $\mathcal{L}_R$ differ only in temperature $($the mean $\bar{n}(\omega))$, then the L and R heat currents are interchangeable when the respective temperatures are interchanged, so that complete heat reciprocity obtains. In order to have non-reciprocity we must introduce asymmetry between the L and R Lindblad superoperators, independently of the respective temperatures. One way to incur such asymmetry is by composing the multilevel system of two subsystems with L-R asymmetric interaction. A minimal model for such a composite system is given in Eq.~(\ref{eq:H1}).
 
The FSD puts severe restrictions on the rectification. This can be seen upon considering the possible channels that contribute to global heat transfer, i.e. heat transfer between the baths in Fig.~\ref{fig:qqRectMechanism}.
Similar energy cycles can be identified for 
OMS, as in the example shown in Figs.~\ref{fig:OM-RA}(a)-(b).
The rectification factor falls short of 1 depending on the coupling strength between the subsystems, the energy mismatch of the subsystems and the bath temperatures. In particular, for symmetric (identical) qubits, the rectification factor is typically much less than 1, as shown in App.~\ref{AppendixC}. However, conditions (Eq.~(\ref{eq:Rspectrum})) ensure that only one bath contributes to the heat flow in either direction, and thus leaves only one unidirectional heat-transfer channel  (Raman cycle) intact in Fig.~\ref{fig:qqRectMechanism}. This ensures perfect rectification, regardless of whether the qubits are identical (resonant) or what the magnitude of $g$ is (Fig.~\ref{fig:qRect}), as detailed in App.~\ref{AppendixC}. By contrast, FSD or regular Lorentzian BSF allow for other, bi-directional channels due to spectral overlap of the two baths and therefore can yield weak rectification. There are, however, situations where skewed-Lorentzian BSF is not mandatory for rectification (App.~\ref{AppendixC}).
\section{HEAT-TRANSISTOR AMPLIFICATION WITH BSF}\label{sec:Ampli}
\begin{figure}[t]
\centering
\includegraphics[width=0.45\textwidth]{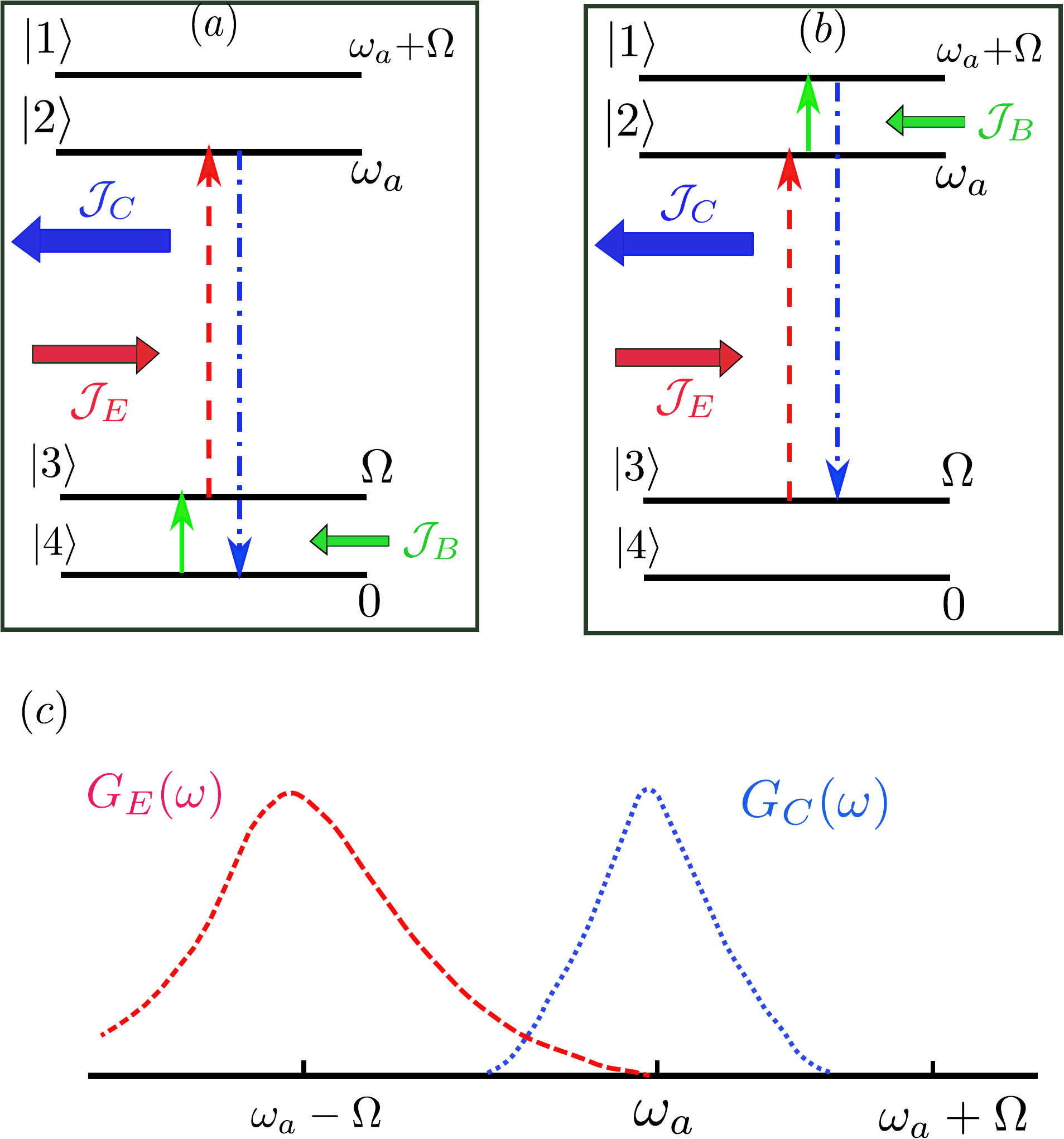}
\caption{(Color online) The processes involved in optimal HT heat transfer via coupled TLS under BSF. The choice of skewed-Lorentzian filtered bath spectra for E and C baths is  shown in panel (c), whereas the B bath has FSD. In panels (a) and (b) the Raman cycles (4324) and (3213), respectively, and their inverse are the only allowed cycles for heat transfer. The dashed,  dashed-dotted and solid arrows represent the transitions induced by the C, E, and B baths, respectively. All other cycles, including four-wave mixing cycles, are prohibited.  The B current is associated with the lowest transition frequency ($\Omega$) in the system. The horizontal arrows denote the heat currents $\mathcal{J}_{E}$, $\mathcal{J}_{C}$ and $\mathcal{J}_{B}$.}
\label{fig:qqAmpMech}
\end{figure} 
A three-terminal heat-transistor (HT)  setup is comprised of three baths dubbed base (B), emitter (E) and collector (C) that are coupled via a controller system S (Fig.~\ref{fig:BareQubits}), where the frequency of subsystem A (B) is labeled by $\omega_{a}$ ($\omega_{b}$).
The HT functions properly if a small change of the B temperature results in a massively amplified heat flow through E or C. The amplification factor is~\cite{PhysRevLett.116.200601, Zhang_2018, PhysRevA.97.052112, PhysRevE.99.032114, PhysRevE.99.042102, PhysRevE.99.062123}
\begin{equation}\label{Eq:amplification}
\alpha_{E, C} = \frac{\partial{\mathcal{J}_{E, C}}}{\partial{\mathcal{J}_{B}}},
\end{equation}
or equivalently, upon using energy conservation, as
\begin{equation}\label{Eq:NDTR}
\alpha_{E}=\Biggl|\frac{\partial(\mathcal{J}_{E})}{\partial(\mathcal{J}_{E}+\mathcal{J}_{C})}\Biggr|= \Biggl|\frac{R_{E}}{R_{E}+R_{C}}\Biggr|.
\end{equation}

Here $R_{E}=(\partial{\mathcal{J}_{E}}/\partial{\mathcal{J}_{B}})^{-1}_{T_{\text{E=const}}}$ and $R_{C}=-(\partial{\mathcal{J}_{C}}/\partial{\mathcal{J}_{B}})^{-1}_{T_{\text{C=const}}}$ are differential thermal resistances. Similar relations can be written for $\alpha_C$. Amplification factors larger than 1 require, according to Eq.~(\ref{Eq:NDTR}),  negative  differential thermal resistance (NDTR), $R_{E}\times R_{C}<0$.

\begin{figure}
\centering
\hspace*{-0.5cm}
\subfloat[]{\includegraphics[width=0.50\textwidth]{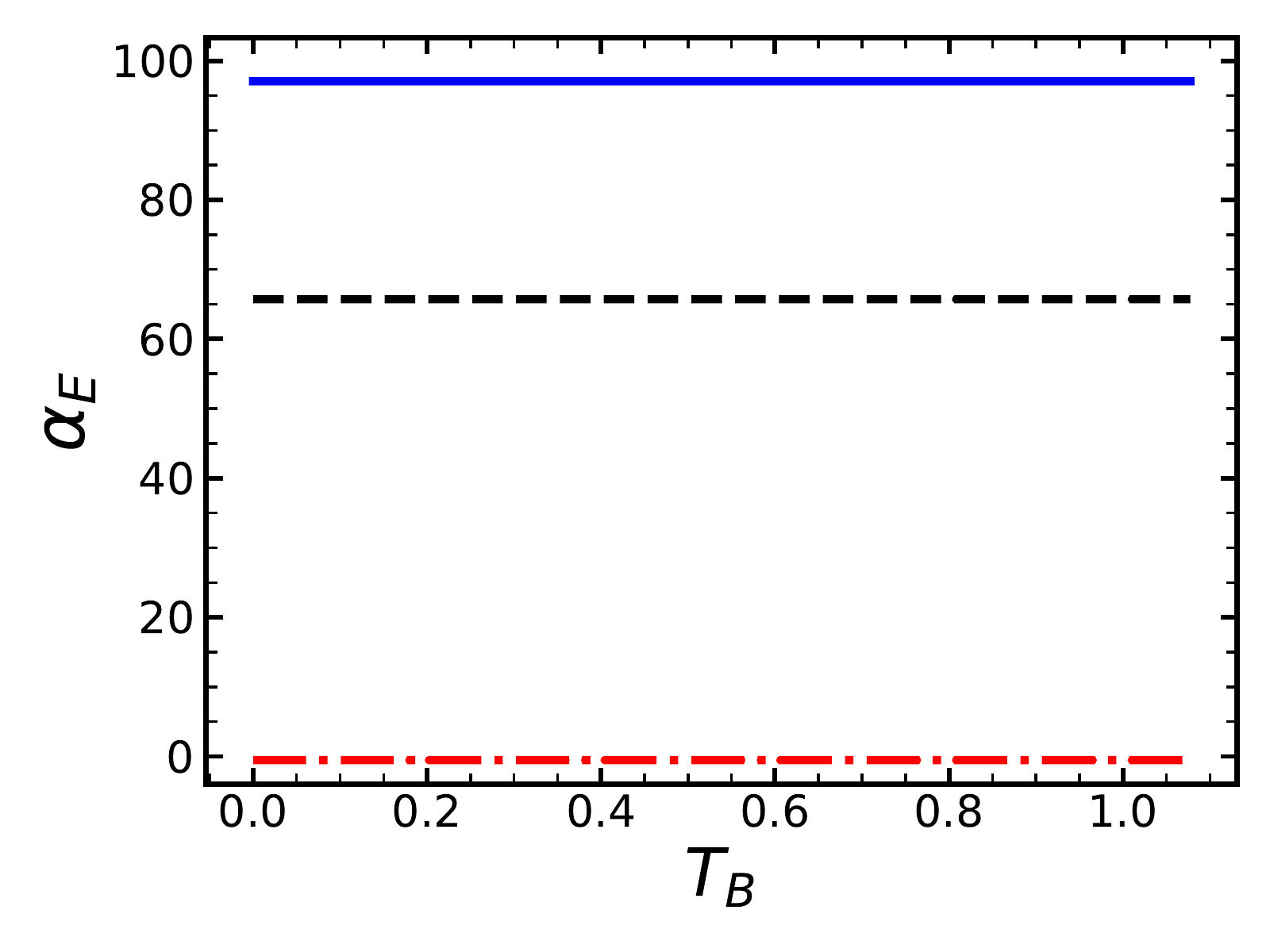}}
\caption{(Color online)~ HT amplification $\alpha_{E}$  as a function of the base temperature $T_{B}$ for coupled TLS. The solid, dashed, and dot-dashed lines are for BSF with the choice of bath spectra shown in Fig.~\ref{fig:qqAmpMech}(c), Fig.~\ref{fig:AppD}(c), and unfiltered baths , respectively. BSF gives rise to dramatic boost in amplification. 
Parameters: $\omega_{a}=1$, $\omega_{b}=0.01$, $g=0.005$, $\kappa_{C}=\kappa_{E}=\kappa_{B}=0.001$, $T_{E}=1$, and $T_{C}=0.01$.}
\label{fig:HeatAmplification}
\end{figure}

The question we pose here is: which mechanisms ensure that the amplification  factors exceed 1  and that  NDTR holds in the minimal models considered above? For two coupled qubits under appropriate BSF only 2 (Raman) cycles and their inverse (Fig.~\ref{fig:qqAmpMech})  are open channels for heat flow between all three baths, whereby energy absorption from B results in energy transfer from E to C (Fig.~\ref{fig:qqAmpMech}(a)) or vice versa (Fig.~\ref{fig:qqAmpMech}(b)). The heat currents are then given by (App.~\ref{AppendixD})
\begin{eqnarray}\label{eq:heatK}
\mathcal{J}_{E} = \omega_{-} K, \qquad
\mathcal{J}_{C} = -\omega_{a} K,\qquad
\mathcal{J}_{B} = \Omega K,
\end{eqnarray}
where, $\omega_{-}:=\omega_{a}-\Omega$, $K= s^2\Gamma_{\text{T}}$, $s=2g/\Omega$.
The expression for the factor  $\Gamma_{\text{T}}$ in the steady-state currents under the chosen BSF (App.~\ref{AppendixD}) shows that $\mathcal{J}_{C}, \mathcal{J}_{E}$ and $\mathcal{J}_{B}$ are associated with the transition frequencies $\omega_a, \omega_{-}$ and $\Omega$, respectively. For weakly coupled qubits, we have $\omega_a > \omega_{-}\gg \Omega$, so that, accordingly, 
$\mathcal{J}_{C},\mathcal{J}_{E}\gg\mathcal{J}_{B}$, and the amplification factors become $\alpha_{E}=\omega_{-}/\Omega$, and $\alpha_{C}=-\omega_{a}/\Omega$.
This choice of the BSF ensures the optimal HT regime (App.~\ref{AppendixD}), wherein small increase in $T_B $ results in  only slight increase of $\mathcal{J}_{B}$, but in large increase of $\mathcal{J}_{E}$  and $\mathcal{J}_{C}$. The heat currents are shown in Figs.~\ref{fig:HeatAmplification} and~\ref{fig:AppD2} for the bath spectral densities in Fig.~\ref{fig:qqAmpMech}(c) and Fig.~\ref{fig:AppD}(c).

Energy cycles similar to those presented in Fig.~\ref{fig:qqAmpMech} describe
heat amplification in the case of OMS, as explained in Fig.~\ref{fig:OM-RA}(c). We note that, for OMS, $\omega_{b}=\Omega$, and Eq.~(\ref{eq:heatK}) only holds in the weak coupling regime $g^2\langle\tilde{b}^{\dagger}\tilde{b}\rangle\ll\omega^{2}_{b}$, where the transformed creation (annhilation) bosonic operators  $\tilde{b}^{\dagger}$($\tilde{b}$) are given in App.~\ref{AppendixA}. In OMS based on optical photons, the photon frequency greatly surpasses the phonon frequency, $\omega_{a}\gg\omega_{b}$, consequently, we can get very large heat amplification. However, for microwave photons the frequency may be close to that of the phonons, leading to suppressed heat amplification.

Our main point  is that the amplification factor, which depends on the two-qubit coupling strength $g$, can be strongly boosted by resorting to BSF, as in the case of rectification. The amplification boost can be understood by noting that in the optimal HT regime (Fig. \ref{fig:qqAmpMech})  we have, from energy conservation, opposite heat flows from E and C, 
\begin{equation}
\abs{\mathcal{J}_{C}}\approx\abs{\mathcal{J}_{E}}.
\end{equation}
To avoid mutual cancelation of these heat flows, which can strongly inhibit the amplification factors $\alpha_{E}$ and  $\alpha_{C}$ (App.~\ref{AppendixD}), we must {\it spectrally isolate} $\mathcal{J}_{C}$ and $\mathcal{J}_{E}$. To this end, we should couple the transition frequencies associated with the E and C heat flows to separate baths: The E bath should couple only to the transition at frequency $\omega_E=\omega_a$, and the C bath only to that at $\omega_C= \omega_a- \Omega$. This condition amounts to the separation of the C and E coupling spectra by BSF, which again corresponds to skewed-Lorentzians with weakly overlapping tails (Fig.~\ref{fig:qqAmpMech}). By contrast, since the B current feeds on the much lower $\Omega$  and is much smaller than the other currents, it is unaffected by BSF:  the B bath may conform to FSD.

The amplification boost due to BSF can be dramatic, as shown in Fig. \ref{fig:HeatAmplification}. Similar effect is obtained for the amplification if the transitions at the frequencies $\Omega$ , $\omega_a$ and $\omega_a+\Omega$  are induced by the B, E, and  C baths respectively. 
 
\section{\label{sec:ExpReal}Experimental realizations}

\begin{figure}
  \centering
  \subfloat[]{\includegraphics[width=0.4\textwidth]{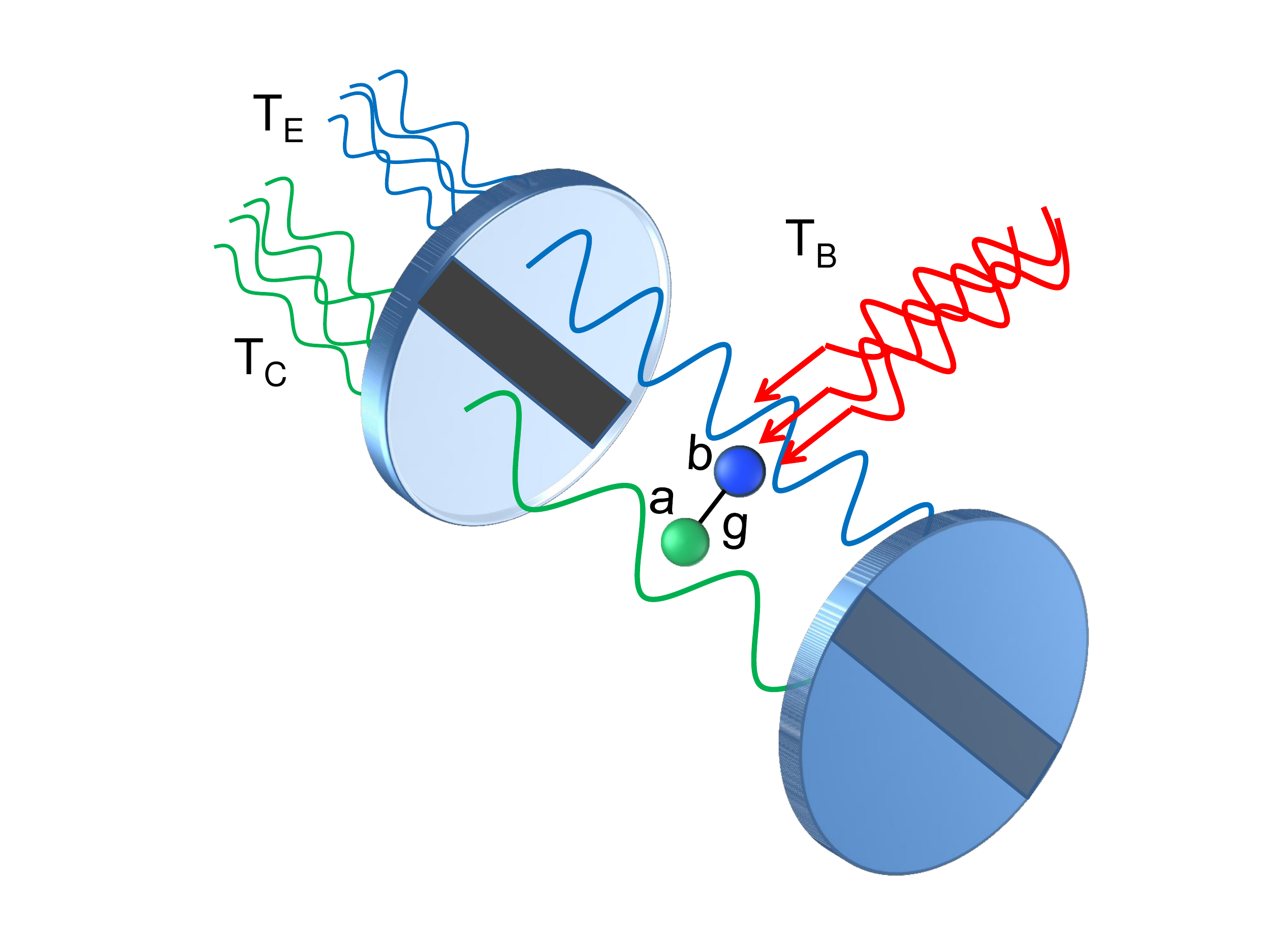}}\\
  \subfloat[]{\includegraphics[width=0.4\textwidth]{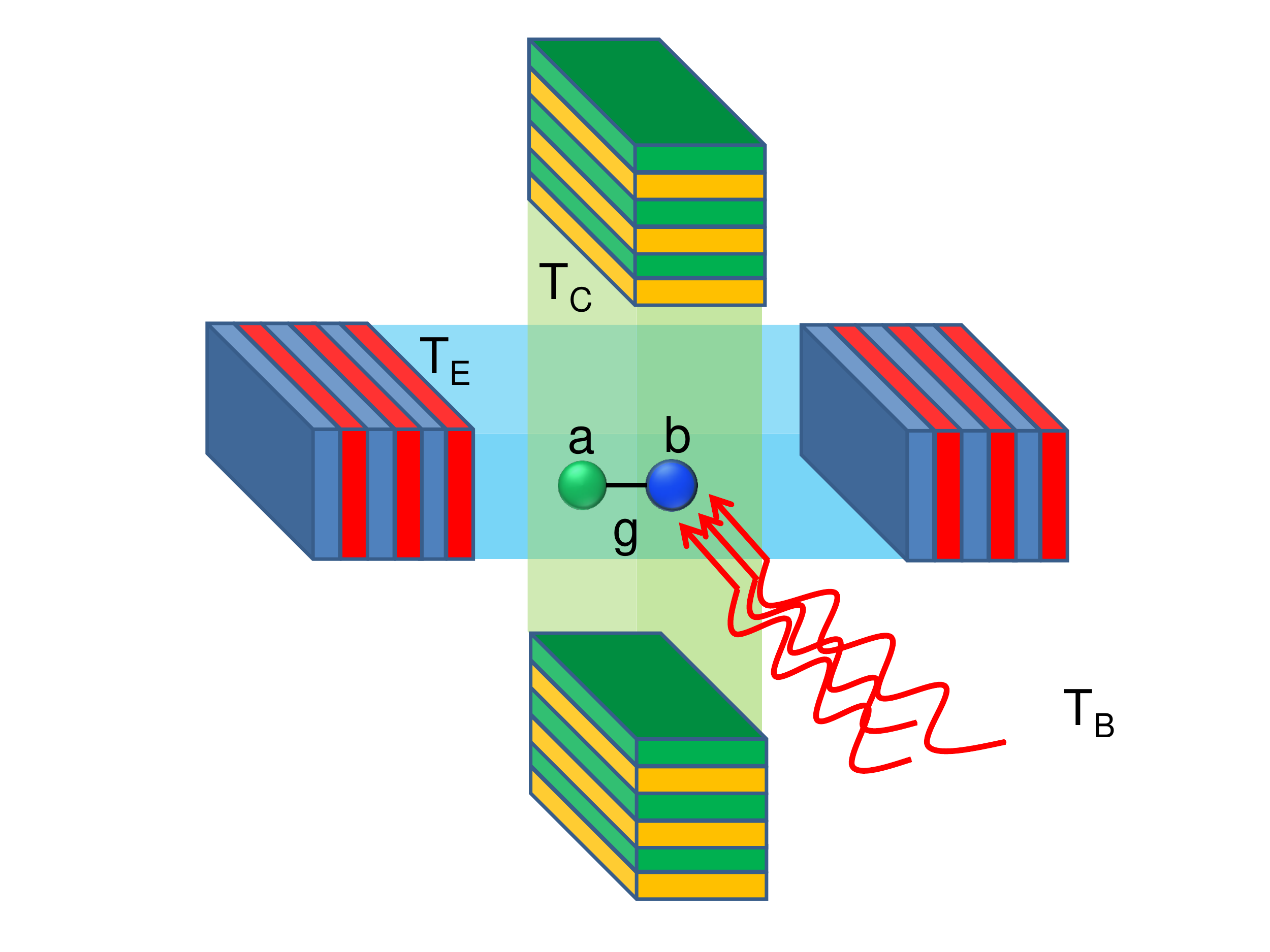}}
  \caption{(a) Realization of the model in Fig.~\ref{fig:BareQubits} for coupled qubits in a bimodal cavity.
  The qubits (a and b) are coupled via dipole-dipole interaction with separation dependent coupling strength $g$ that can be engineered in a cavity. The qubits are coupled to the cavity modes that act as BSF: filters coupling the qubit a (b) to the thermal bath at temperature $T_{C}$ ($T_{E}$). A heat insulating strip (black) allows the coexistence of two regions with different temperatures. For HT configuration, the qubits are exposed to thermal radiations at a much lower frequency that plays the role of base bath B at temperature $T_{b}$. (b) Realization of the same model with skewed-Lorentzian BSF inside distributed Bragg reflectors (DBR). The coupled qubits (a and b) are placed at the crossing region of the two cavity fields.}
 \label{fig:concul}
\end{figure}

The advantageous HD or HT schemes boosted by BSF described here can be experimentally realized in a number of setups that may employ analogs of the quantum optical methods proposed here and in Ref.~\cite{PhysRevE.99.042121}:

A.  Solid-state setups, may be based on qubits that are  NV centers or similar defects with optical and microwave transitions. These qubits can be embedded in a bimodal cavity. Each qubit can be near-resonantly coupled to another cavity mode that has an antinode  at its location. That cavity mode acts as a filter coupling the qubit to one or another thermal bath whose temperature is set by the respective pair of mirrors, as shown in Fig.~\ref{fig:concul}(a). Two-qubit coupling can be mediated by their  near-resonant dipole-dipole interaction~\cite{PhysRevLett.89.207902} whose separation- dependent strength may be drastically modified in cavities~\cite{PhysRevA.53.R35}, waveguides~\cite{PhysRevA.83.033806,Shahmoon10485}  or in periodic structures with photonic band gaps~\cite{PhysRevA.42.2915, lambropoulos_fundamental_2000,Shahmoon:16, liu_quantum_2017}. Heat-flow rectification can be observed upon interchanging the mirror temperatures. When the cavity finesse is high enough,  BSF takes place, the bath spectra associated with the two high-Q modes have suppressed overlap, and rectification is boosted. A HT configuration can be realized by exposing all qubits to the same thermal radiation at a much lower (microwave or far-infrared) frequency that plays the role of the base bath B (Fig.~\ref{fig:concul}(a)). This configuration can exhibit skewed-Lorentzian BSF when the qubits are placed within distributed Bragg reflectors (DBR), as shown in the Fig.~\ref{fig:concul}(b).

B. Optomechanical systems with spectrally structured phononic baths engineered by sonic DBR exhibit similar BSF. They can rely on recent progress in phononic bandgap materials and phononic cavities~\cite{khelif_phononic_2015, deymier_acoustic_2013,gorishnyy_sound_2005,
kushwaha_acoustic_1993,khelif_transmission_2003,
ho_broadband_2003,sigalas_defect_1998,
elnady_quenching_2009,khelif_guiding_2004,tian_rainbow_2017}.

C. Electronic circuits composed of transmon qubits in superconducting cavities~\cite{Pekola2015} have been identified as possible HD~\cite{senior2019heat}. The analog of a mechanical mode in such circuits may be implemented by a transmission line and optomechanical-like coupling 
 can be induced between superconducting microwave resonators~\cite{PhysRevA.90.053833, PhysRevLett.120.227702}.\\

\section{Conclusions}\label{sec:Conclu}

The analysis presented here has underscored the ability of bath spectral filtering (BSF) to serve as a resource for boosting the performance of a multifunctional heat manager, particularly as heat diode (HD) or heat transistor (HT), when this manager is based on a quantum  heat-control system with the minimal number of degrees of freedom.
The key to optimize HD performance, i.e. achieve perfect heat-flow rectification, is asymmetry in the coupling of the control system to left-and right-hand baths. Such asymmetric coupling between left and right hand subsystems can be incurred via the coupling of two subsystems whose minimal models are anisotropic two-qubit coupling or optomechanical system (OMS) coupling. However, restrictions on the coupling strength imply that adequate asymmetry is not always attainable, particularly for identical qubits or resonant OMS. 

Remarkably, BSF that yield {\it strongly skewed-Lorentzian bath lineshapes}  has been shown here to enable perfect rectification of an HD regardless of such restrictions.  Equally remarkable is the finding that skewed-Lorentzian BSF applied to the collector and emitter baths can boost HT amplification in the optimal regime where the  base heat current is very small, so that the collector and emitter heat currents are nearly equal in magnitude and flow in opposite directions. For suitable system parameters, the proposed setup in Fig.~{\ref{fig:BareQubits}} can also be employed as a thermal switch~\cite{Karimi_2017} or a heat valve~\cite{Ronzani2018}, and these functionalities may also be strongly boosted by skewed-Lorentzian BSF. Essentially, the beneficial role of such BSF in quantum heat management is akin to that previously predicted for qubit-based minimal heat engines and refrigerators~\cite{Gelbwaser_2015, PhysRevE.90.022102, Gelbwaser-Klimovsky2015, PhysRevE.87.012140, Ghosh2018, arXiv:2002.11472} that must be coupled to spectrally-separated hot and cold baths in order to attain high efficiency or power. 

The guidelines for engineering the skewed-Lorentzian BSF in Eq. (11) to ensure HD and HT performance boost (as in Eq. (14)) are essentially as follows: Introduce a cut-off or at least sharp drop-off on the required spectral wing of the Lorentzian by appropriately selecting the filter-mode frequency $\tilde\omega_\alpha$ and the Lamb shift in Eq. (12). In general, distributed Bragg reflectors (DBR) possess band gaps that allow for such engineering \cite{JMO94,Shahmoon10485,PhysRevA.42.2915, lambropoulos_fundamental_2000,Shahmoon:16, liu_quantum_2017}.

The predicted BSF boost of quantum heat management may open a new avenue in our ability to pursue advantageous thermodynamic functionalities based on quantum systems, by exploiting hitherto untapped resources. As opposed to prevailing resources employed in  quantum HT and HD schemes that have classical analogs, the BSF is a uniquely quantum mechanical resource:  it stems from the renormalization of the interaction between the system and the bath and is manifest in their dissipative rate of energy exchange as well as in the bath-induced dispersive (Lamb ) shift~\cite{JMO94,lambropoulos_fundamental_2000} (see App.~\ref{AppendixB} )\\

{\bf Acknowledgements:} G.K acknowledges the support of ISF, DFG, SAERI, PATHOS (EU) and PACE-IN (QUANTERA).

\appendix
\section{\label{AppendixA} Master Equation Derivation }
\label{app:MasterDeri}
Here we present the master equation for the cases considered in the main text. The Hamiltonian for the three independent baths is given by
\begin{equation}
\hat{H}_{\alpha}=\sum_{k}\omega_{k}\hat{a}^{\dagger\alpha}_{k}\hat{a}^{\alpha}_{k},
\end{equation}
where $\hat{a}^{\dagger\alpha}_{k}$ ($\hat{a}^{\alpha}_{k}$) are the creation (annihilation) operator of the $k$'th mode of the bath $\alpha = E, C, B$. The system bath interaction Hamiltonian has the form
\begin{eqnarray}\label{eq:sysbath}
H_{SB}=\hat{s}_{a}\otimes\sum_{k}g^{i}_{k}(\hat{a}^{i}_{k}+\hat{a}^{\dagger i}_{k})+\hat{s}_{b}\otimes\sum_{k}g^{B}_{k}(\hat{a}^{B}_{k}+\hat{a}^{\dagger B}_{k})\nonumber\\
\end{eqnarray}
where $i=E, C$ and the $\hat{s}_{a}$ ($\hat{s}_{b}$) annihilation operator pertains to the right (left) sub-system. In Sec.~\ref{sec:Rect}, 
for the discussion of rectification, subsystem a (b) is labeled by L (R) and the bath temperature $T_{E}$
 ($T_{B}$) is relabeled as $T_{L}$ ($T_{R}$).
\\
 We next consider the cases of two coupled two-level systems( TLS or qubits) and two coupled harmonic oscillators:
(i) For the coupled TLS,  $\hat{s}_{a, b}$ are $\hat{\sigma}^{x}_{a, b}$ Pauli operators in Eq.~(\ref{eq:sysbath}). The system Hamiltonian given in Eq.~(\ref{eq:H1}), can be diagonalized by the unitary transformation
\begin{equation}\label{eq:qqTransform}
U := \exp\Big(-i\frac{\theta}{2}\hat{\sigma}_{a}^{z}\hat{\sigma}_{b}^{y}\Big),
\end{equation}
where the angle $\theta$ is defined as sin $\theta:=2g/\Omega$ and cos $\theta:=\omega_{b}/\Omega$ such that $\Omega := \sqrt{\omega^{2}_{b}+4g^2}$. The diagonalized Hamiltonian takes the form
\begin{equation}
\tilde{H}_{qq} = \frac{\omega_{a}}{2}\tilde{\sigma}^{z}_{a} + \frac{\Omega}{2}\tilde{\sigma}^{z}_{b},
\end{equation}
and the transformed operators read
\begin{eqnarray}
\tilde{\sigma}^{z}_{a} &=&\hat{\sigma}^{z}_{a}, \\
\tilde{\sigma}^{z}_{b} &=&\text{cos} \theta\hat{\sigma}^{z}_{a}+ \text{sin}\theta\hat{\sigma}^{z}_{a}\hat{\sigma}^{x}_{b}.
\end{eqnarray}
The eigenstates of the diagonalized Hamiltonian are represented by $\ket{j}$, $j=1,2,3,4$ with their corresponding eigenvalues $\omega_{1} = \frac{1}{2}(\omega_{a} + \Omega)$,  $\omega_{2} = \frac{1}{2}(\omega_{a} - \Omega)$, $\omega_{3} = \frac{1}{2}(-\omega_{a} + \Omega)$, and $\omega_{4} = \frac{1}{2}(-\omega_{a} - \Omega)$, respectively.
In order to derive the master equation, we transform the operators to the interaction picture in which
\begin{eqnarray}\label{eq:tdependeqs}
\hat{\sigma}_{a}^{x}(t)&=&\cos\theta{\tilde{\sigma}}_{a}^{-}e^{-i\omega_{a}t}-\sin\theta{\tilde{\sigma}}_{{a}}^{+}{\tilde{\sigma}}_{{b}}^{-}e^{-i(\Omega-\omega_{a})t} \\\nonumber&&+\sin\theta{\tilde{\sigma}}_{{a}}^{-}{\tilde{\sigma}}_{{b}}^{-}e^{-i(\Omega+\omega_{a})t} + \mathrm{H.c.}\\
\hat{\sigma}_{{b}}^{x}(t)&=&\cos\theta{\tilde{\sigma}}_{{b}}^{-}e^{-i\Omega t} + \frac{1}{2}\sin\theta{\tilde{\sigma}}_{{a}}^{z}{\tilde{\sigma}}_{{b}}^{z} + \mathrm{H.c.}
\end{eqnarray}
The master equation in the interaction picture evaluates to 
\begin{eqnarray}\label{eq:master}
&\dot{\hat{\rho}}& = \left(\hat{\mathcal{L}}_{E} +\hat{\mathcal{L}}_{C} +\hat{ \mathcal{L}}_{B}\right)(\hat{\rho}),
\end{eqnarray}
where the superoperators $\hat{\mathcal{L}}_{\alpha}\hat{\rho}$ with $\alpha= E, C, B$, describing the quantum dynamics of the system that interacts with the baths, are of the form~\cite{PhysRevE.99.042121} 
\begin{eqnarray}
\label{eq:AppL_L}
\hat{\mathcal{L}}_{o, i} &=& G_{i}(\omega_{a})c^{2}\hat{\mathcal{D}}[\tilde{\sigma}^{-}_{a}]
+ G_{i}(-\omega_{a})c^{2}\hat{\mathcal{D}}[\tilde{\sigma}^{+}_{a}], \\\nonumber
\hat{\mathcal{L}}_{q, i} &=& G_{i}(\omega_{q})s^{2}\hat{\mathcal{D}}[\tilde{w}_{q}]
+ G_{i}(-\omega_{q})s^{2}\hat{\mathcal{D}}[\tilde{w}^{\dagger}_{q}], q=\pm 1
\\
\hat{ \mathcal{L}}_{B}&=& G_{B}(\Omega)c^{2}\hat{\mathcal{D}}[\tilde{\sigma}^{-}_{b}]
+ G_{B}(-\Omega)c^{2}\hat{\mathcal{D}}[\tilde{\sigma}^{+}_{b}], \label{eq:AppL_R}
\end{eqnarray}
where $\hat{\mathcal{L}}_{i}=\hat{\mathcal{L}}_{o, i}+\hat{\mathcal{L}}_{q, i}$, $i=E, C$, $\tilde{w}_{1}=\tilde{\sigma}^{-}_{a}\tilde{\sigma}^{-}_{b}$, $\tilde{w}_{-1}=\tilde{\sigma}^{-}_{a}\tilde{\sigma}^{+}_{b}$, $\omega_{\pm 1}=\omega_{a}\pm\Omega$, $c=\cos\theta$, and $s=\sin\theta$.\\
\begin{figure}[t!]
\centering
\hspace*{-0.5cm}
\includegraphics[scale=0.27]{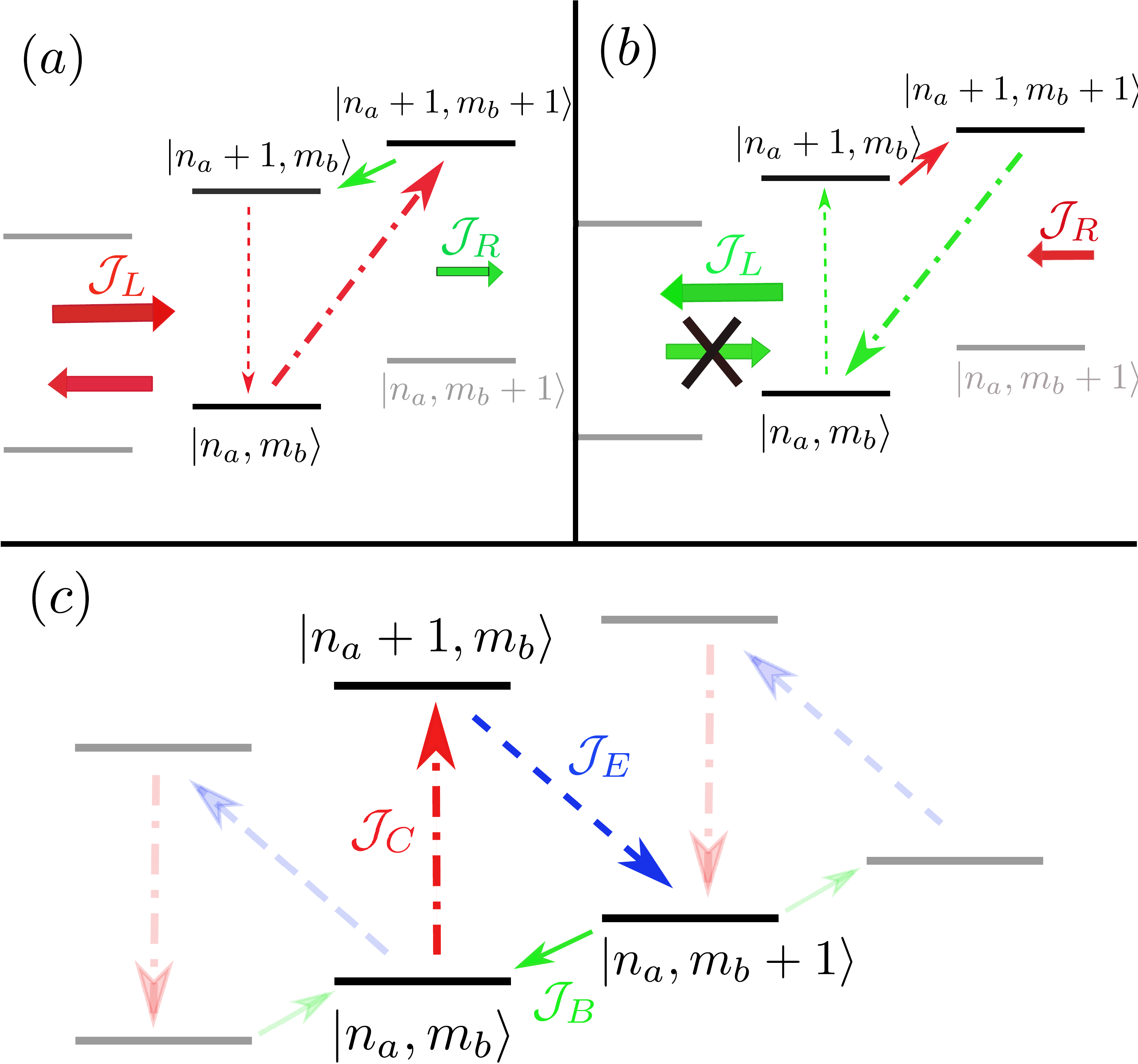}
\caption{(color online) (a), (b) Examples of the processes involved in heat transfer via an OMS. In the eigenstates $\ket{n_{a}, m_{b}}$, $n_{a}=0,1,2,...$  corresponds to the number of photons, and $m_{b}$ to the number of phonons in the $n$th photon sub-space.  As for coupled TLS, the cold bath may not be able to excite $\ket{n_{a}, m_{b}+1}\to\ket{n_{a}+1, m_{b}+1}$ transition shown in panel (b), resulting in the rectification of heat flow from left to right. Panel (c) shows examples of processes that transfer heat between all three baths. The base heat current $\mathcal{J}_{B}$ is associated with the lowest  transition frequency in the system.}
\label{fig:OM-RA}
\end{figure}
(ii) For the optomechanical system (OMS), $\hat{s}_{a}=(\hat{a}+\hat{a}^{\dagger})$, $\hat{s}_{b}=(\hat{b}+\hat{b}^{\dagger})$, the unitary transformation
\begin{equation}\label{eq:omTransform}
U_{om} := \exp\Big(\beta\hat{a}^{\dagger}\hat{a}(\hat{b}^{\dagger}-\hat{b})\Big),
\end{equation}
diagonalizes the optomechanical Hamiltonian and takes the form
\begin{equation}\label{eq:omDiag}
\tilde{H}_{om} = \omega_{a}\tilde{a}^{\dagger}\tilde{a}+\omega_{b}\tilde{b}^{\dagger}\tilde{b}-\frac{g^2}{\omega_{b}}(\tilde{a}^{\dagger}\tilde{a})^2.
\end{equation}  
The eigenstates of this Hamiltonian are $\ket{n_{a}, m_{b}}$, where $m_{b}$ are the number of phonons in the $n_{a}$th photon-subpace. The transformed operators in the interaction picture are given by
\begin{eqnarray}\label{eq:approxQO}
\hat{a}(t) &=&\tilde{a}e^{-i\omega_{a}t}\sum^{\infty}_{n=0}\beta^{n}(\tilde{b}e^{-i\omega_{b}t}-\tilde{b}^{\dagger}e^{i\omega_{b}t})^{n}\nonumber\\
&\approx&\tilde{a} e^{-i\omega_{a}t} + \beta\tilde{a}\tilde{b}e^{-i(\omega_{a}+\omega_{b})t} - \beta\tilde{a}\tilde{b}^{\dagger}e^{-i(\omega_{a}-\omega_{b})t},\nonumber\\
\hat{b}(t) &=&\tilde{b}e^{-i\omega_{b}t}+\beta\tilde{a}^{\dagger}\tilde{a}.
\end{eqnarray} 
The master equation in the interaction picture is the one given in  Eq.~(\ref{eq:master}). For the OMS, the dissipators in Eqs.~(\ref{eq:AppL_L}) and (\ref{eq:AppL_R}) have $c=1$, $s=g/\omega_{b}$, and the Pauli operators $\tilde{\sigma}^{-}_{a}$ ($\tilde{\sigma}^{-}_{b}$) are replaced by the bosonic operators $\tilde{a}$ ($\tilde{b}$).

\section{\label{AppendixB} BSF for OMS}
Generically, an optomechanical-like interaction between two subsystems labeled with A and B is of the form
\begin{eqnarray}
H=\omega_a a^\dag a+\omega_b b^\dag b+ga^\dag a(b+b^\dag),
\end{eqnarray}
where $a,b$ and $a^\dag,b^\dag$ are the annihilation and creation ladder operators of excitations in corresponding subsystems. The excitation frequencies are denoted by $\omega_a,\omega_b$ and $g$ stands for the coupling coefficient.
Subsystem $A$ consists of either spins or bosons , while subsystem $B$ is assumed to be always bosonic. Both subsystems are attached to their respective thermal reservoirs, which are modeled as ensembles of harmonic oscillators described by
\begin{eqnarray}
H_{a-\text{bath}}&=&\sum_q\eta_a(q) (a^\dag a_q+a a_q^\dag),\\
H_{b-\text{bath}}&=&\sum_q\eta_b(q) (b^\dag b_q+b b_q^\dag),
\end{eqnarray}
where the rotating wave approximation is employed and the system- bath coupling strengths $\eta_a,\eta_b$ are assumed to be weak relative to $\omega_a,\omega_b$. 
After a canonical transformation (similar to the Fr\"{o}lich polaron or the Schrieffer-Wolff transformation) with the unitary $U=\exp{(S)}$ where
\begin{eqnarray}
S=\frac{g}{\omega_b}a^\dag a(b^\dag-b)
\end{eqnarray}
the system Hamiltonian changes to $H'=U^\dag H U$ and reads
\begin{eqnarray}
H'=\omega_a a^\dag a+\chi (a^\dag a)^2+\omega_b b^\dag b
\end{eqnarray}
and the system-bath interactions become
\begin{eqnarray}
H'_{a-\text{bath}}&=&\sum_q\eta_a(q) (a^\dag\text{e}^{\bar{g}(b^\dag-b)} a_q+h.c.),\\
H'_{b-\text{bath}}&=&\sum_q\eta_b(q) (b^\dag b_q+b b_q^\dag)
\nonumber\\
&+&\sum_q\bar{g}\eta_b(q) a^\dag a(b_q+b_q^\dag),
\end{eqnarray}
where $\bar g=g/\omega_b$ is introduced for brevity.
For simplicity, we here assume $\omega_{b}\gg g$ so that we can ignore sideband contributions in the system-bath interactions. In addition,
we will neglect the Kerr-type squeezing term in the 
transformed system Hamiltonian. The spectral response function of the bosonic bath coupled to a bosonic system, when the bath is spectrally structured, e.g. has band gaps, can be calculated analogously to that of
a two-level atom in such a structured bath,e.g. in
a photonic band gap material, where a skewed-Lorentzian response function has been found~\cite{Shahmoon10485}. 
For a harmonic oscillator coupled to a bosonic bath,such
studies are conspicuously absent,with one notable exception~\cite{cresser_resolvent_1980}.
We here sketch a simplified derivation
of the spectral response in the case of a fully bosonic system-bath interaction
using the resolvent operator method.
 The resolvent operator, or Green function, method has been developed for sequential decay~\cite{mower_decay_1966,mower_sequential_1968,goldhaber_theory_1967}, and applied to quantum optical systems~\cite{noauthor_quantum_1965,agarwal_quantum_1974,dixit_photon_1980}. It has been  subsequently generalized to structured environments subsequently~\cite{JMO94}. Let us first write the Hamiltonian of our isolated, reduced problem as
\begin{eqnarray}
H=H_0+V,
\end{eqnarray}
where
\begin{eqnarray}
H_0&=&\sum_n\omega (n+1/2)
|n\rangle\langle n|
+\sum_q\omega_q(b_q^\dag b_q),\\
V&=&\sum_{nq}\eta_{nq}(|n+1\rangle\langle n|b_q+|n\rangle\langle n+1|b_q^\dag).
\end{eqnarray}
We look for the broadening and shift of each
energy level of the quantum oscillator system. For that
we assume the system is prepared in the initial state
\begin{eqnarray}
|I\rangle=|n\rangle|0_q\rangle=|n,0_q\rangle,
\end{eqnarray}
denoting the vacuum state of the bath modes as $|0_q\rangle$. The interaction $V$ can couple the initial state to the state 
\begin{eqnarray}
|B_q\rangle=|n-1\rangle|1_q\rangle=|n-1,1_q\rangle.
\end{eqnarray}
The eigenvalues of the states $|I\rangle$ and $|B_q\rangle$ are given by
\begin{eqnarray}
\omega_I&=&\omega(n+1/2),\\
\omega_{B_q}&=&\omega(n-1/2)+\omega_q.
\end{eqnarray}
The equation for the resolvent operator $R(z)$ is given
by~\cite{lambropoulos_fundamental_2000}
\begin{eqnarray}\label{eq:resolvent}
(z-H_0)R(z)=I+VR(z),
\end{eqnarray}
where $I$ is the unit operator.
Taking the matrix element of Eq.~(\ref{eq:resolvent})
$\langle 0_q,n|R(z)|n,0_q\rangle=R_{n;n}$ we find
\begin{eqnarray}
[z-\omega (n+1/2)]R_{n;n}&=&1
+\sum_q\eta_{n-1,q}R_{n-1,1_q;n}\nonumber\\
\end{eqnarray}
This suggests that we need an equation for the matrix element $R_{n-1,1_q;n}$, too. Similarly then we find
\begin{eqnarray}\label{eq:resolvent2}
[z-\omega (n-1/2)-\omega_q]R_{n-1,1_q;n}&=&
\eta_{n-1,q}R_{n;n}\nonumber\\
&+&\eta_{n-2,q}R_{n-1,1_q;n-2,2_q}.\nonumber\\
\end{eqnarray}
The last term can be used for improving  the perturbative
expansion by systematically iterating the recursive relation. Here we will be content with the lowest order
expression and drop the second order terms in interaction strength. Substituting $R_{n-1,1_q;n}$ from Eq.~(\ref{eq:resolvent2}) into Eq.~(\ref{eq:resolvent})
we get
\begin{eqnarray}\label{eq:Gnn}
R_{n;n}=\frac{1}{z-\omega(n+1/2)
-W_n(z)}
\end{eqnarray}
where the so-called shift-width function is
identified to be
\begin{eqnarray}\label{eq:W(z)}
W_n(z)=\sum_q\frac{\eta_{n-1,q}^2}{z-\omega(n-1/2)-\omega_q}.
\end{eqnarray}
At $z=\omega(n+1/2)+i\epsilon$,
the shape is Lorentzian
\begin{eqnarray}\label{eq:W_0}
W_n^{(0)}=\sum_q\frac{\eta_{n-1,q}^2}{\omega-\omega_q+i\epsilon}.
\end{eqnarray}
This expression has a similar form to that obtained for a two-level atom embedded in a bosonic bath~\cite{Shahmoon10485}. Calculating $W(z)$ in the complex-z place by
contour integration with the residue theorem (principal value integration) over a contour just above
the real axis ($-\infty,\infty$) one finds the real part of the response, which is responsible for the bath-induced Lamb shift of the $n$-th level, $\Delta_n(\omega)$. The corresponding imaginary part of the response  yields the $n$-th level width $\Gamma_n$. Due to the similarity with the two-level atom calculation~\cite{Shahmoon10485} we will  only  present the key results.
\begin{figure}[t]
\centering
\hspace*{-0.60cm}
\includegraphics[width=0.5\textwidth]{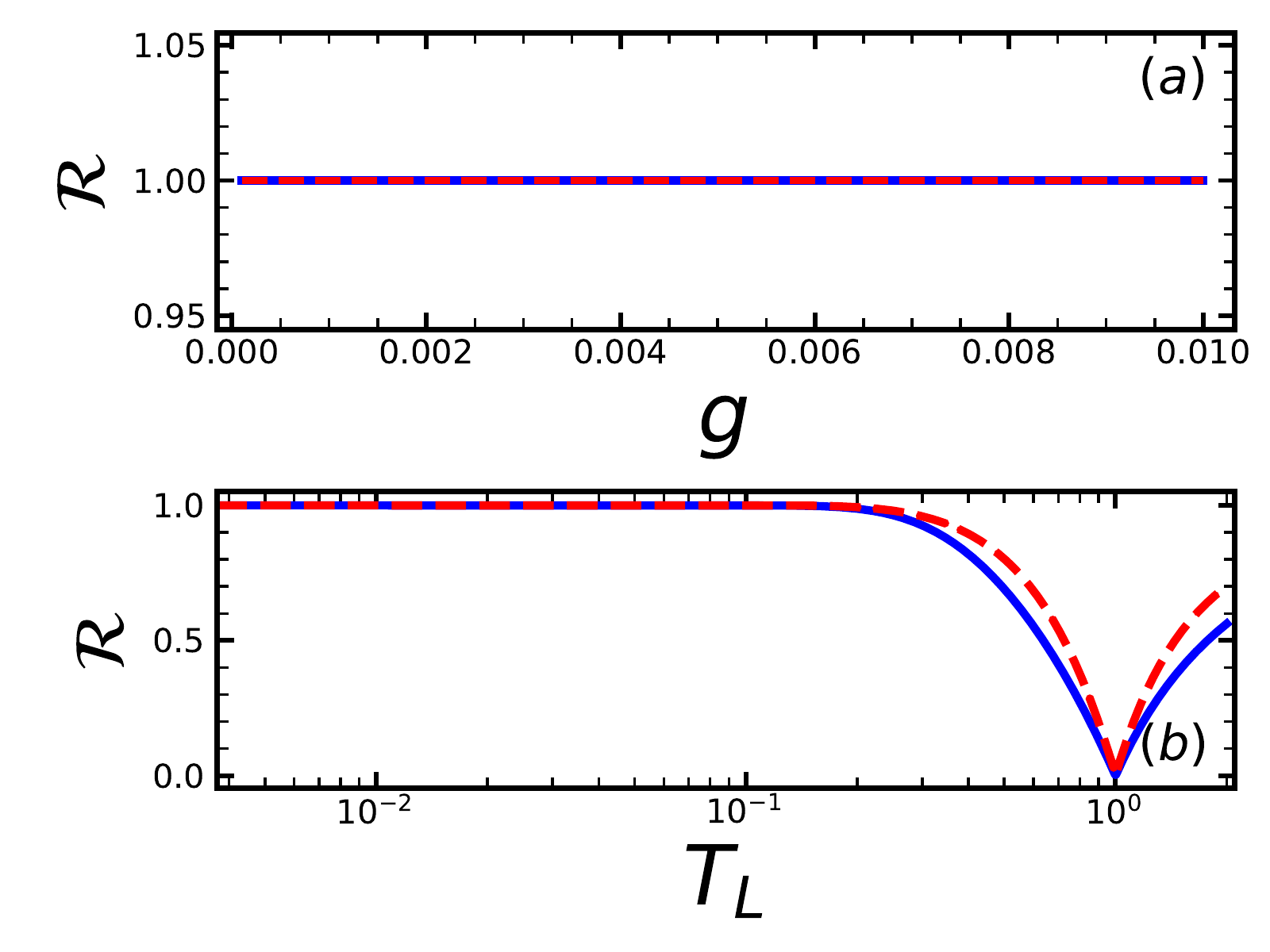}
\caption{(Color online)~{\it
{Rectification}} $\mathcal{R}$ in the HD of~\ref{fig:qqRectMechanism}, (a) as a function of the coupling strength $g$,  (b) as a function of the left bath temperature $T_{L}$. The solid and dashed lines are for coupled TLS, and OMS, respectively. The left bath has filtered spectral density as shown in Fig.~\ref{fig:qqRectMechanism}(c), and the right baths has FSD. In (a) BSF ensures perfect rectification for any g. In (b) rectification is perfect unless temperature gradient is small.
Parameters: $\omega_{L}=1$, $\omega_{R}=0.1$, $\kappa_{L}=\kappa_{R}=0.001$, (a) $T_{L}=2$, $T_{R}=0.02$, and (b) $g=0.01$,  $T_{R}=1$. All the system parameters are scaled with $\omega_{L}/2\pi=10$ GHz.}
\label{fig:AppC}
\end{figure}
 The  spectrally structured character of 
the environment (bath) enters the expression through the density of states $D(\omega_q)$ and the dispersion relation of $\omega_q$,
when the summation over bath modes is replaced by an integral under continuum approximation.
\begin{eqnarray}\label{eq:W(z)Int}
W_{n}(z)=\int\,\text{d}\omega_q\frac{G_{n}(\omega_q)}{z-\omega(n-1/2)-\omega_q},
\end{eqnarray} 
Here
\begin{eqnarray}\label{eq:Sn}
G_{n}(\omega_q)=\sum_\sigma\int d\Omega_q D(\omega_q)\eta_{n-1,q}^2,
\end{eqnarray}
is the spectral response function of the phonon bath. 
The solid angle for the quasimomentum $q$ is denoted by $\Omega_q$. Substituting $W_n^0(\omega)=\Delta_n(\omega)+iG_n(\omega)$ into $R_{n;n}$ and taking the imaginary part yields the quantum(photon or phonon) emission probability, or the bath spectrum, as
\begin{eqnarray}
\tilde{G}_n := \frac{1}{\pi} \frac{\pi G_n(\omega)}{[(\omega - (\omega(n+1/2)+\Delta_n(\omega))^2
+(\pi G_n(\omega))^2]}\nonumber\\,
\end{eqnarray}
where $1/\pi$ is introduced for a skewed-Lorentzian function expression. We may conclude that both photonic or phononic spectrally structured baths would act as a bandpass filter to yield a skewed-Lorentzian for a harmonic oscillator system, similarly to the case of a two-level atom~\cite{PhysRevE.90.022102}.
\section{\label{AppendixC} Rectification with BSF}
\label{recti}
\begin{figure}[t]
\centering
\hspace*{-0.60cm}
\includegraphics[width=0.35\textwidth]{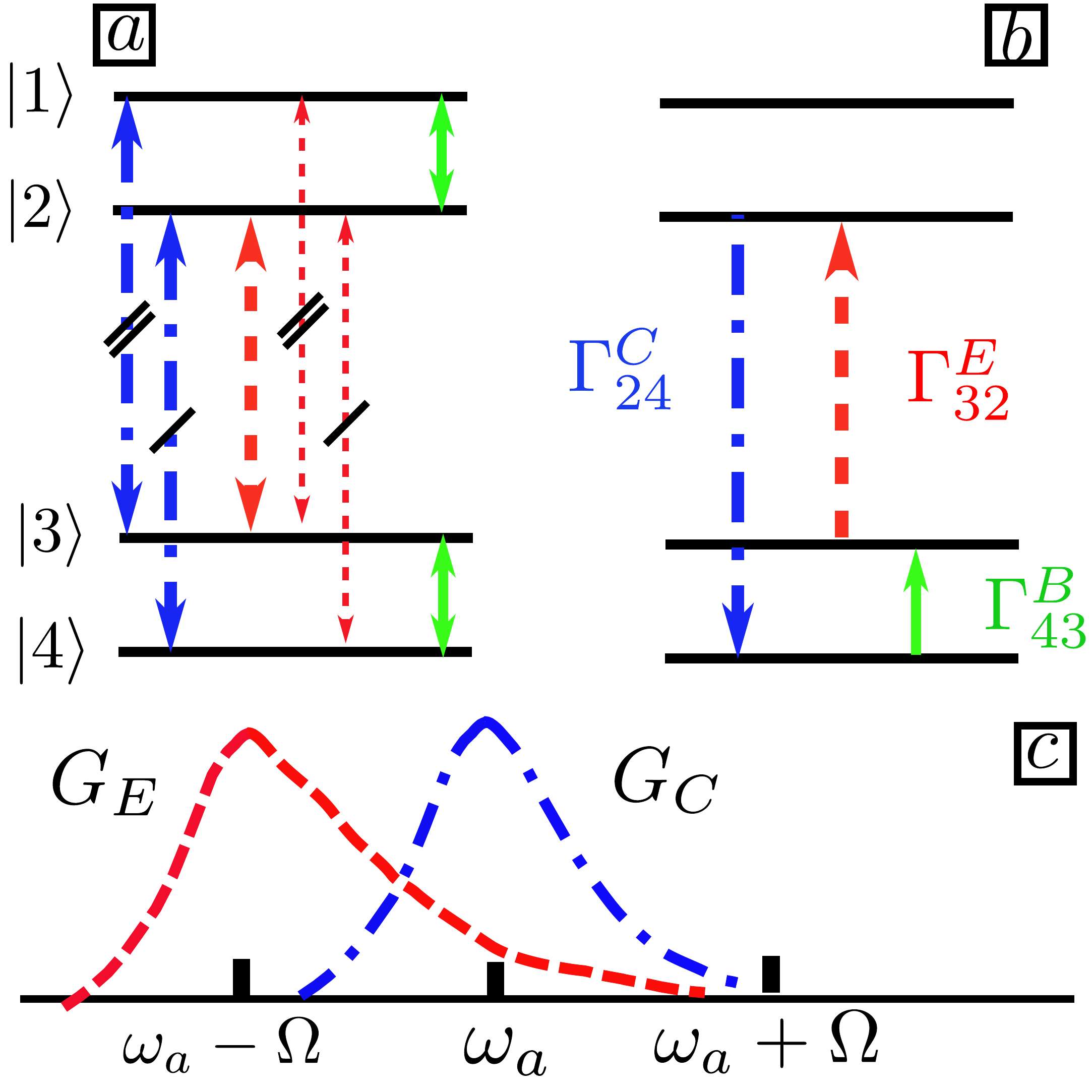}
\caption{(Color online)~(a) The possible transitions induced by the three baths under  suitable BSF. The filtered bath spectra  are presented in (c). The solid, dashed, dot-dashed transitions are induced by the base, emitter and collector baths, respectively. The slashed lines represent direct transfer of heat between the emitter and the collector without passing through the base. (b) An example of an energy cycle that transfers the heat between the three baths. It shows that $\Gamma^{E}_{32}$ and $\Gamma^{E}_{24}$ have opposite signs.}
\label{fig:AppD}
\end{figure}
For the choice of bath spectra shown in Fig.~\ref{fig:qqRectMechanism}(c), the master equation for the coupled TLS in the dressed state basis $\ket{j}$ is given by,
\begin{eqnarray}
\mathcal{L}_{L}(\rho) &=& s^2\Gamma^{L}_{14}(\ket{4}\bra{4}-\ket{1}\bra{1})+ c^2\Gamma^{L}_{24}(\ket{4}\bra{4}-\ket{2}\bra{2}), \nonumber\\
\mathcal{L}_{R}(\rho) &=&  c^2\Gamma^{R}_{12}(\ket{2}\bra{2}-\ket{1}\bra{1}).
\end{eqnarray}
Here, $c=\omega_{R}/\Omega$, $s=2g/\Omega$, and
\begin{equation}
\Gamma^{\alpha}_{ij} = \kappa_{\alpha}\omega_{ij}[(1+\bar{n}_{\alpha}(\omega_{ij}))\rho_{ii}-\bar{n}_{\alpha}(\omega_{ij})\rho_{jj}],
\end{equation}
where $\bar{n}_{\alpha}$ is defined in Eq.~(\ref{eq:SRF}). The steady-state heat current can be evaluated by noting, $\mathcal{L}_{L}+\mathcal{L}_{R}=0$, and $\Gamma^{\alpha}_{ij}=-\Gamma^{\alpha}_{ji}$, and given by,
\begin{eqnarray}\label{eq:heatRect}
\mathcal{J}_{L} = -\Omega c^2 \Gamma , \qquad
\mathcal{J}_{R} = \Omega c^2 \Gamma ,
\end{eqnarray}
where, $\Gamma = \Gamma^{L}_{14} = \Gamma^{R}_{21}$. The expression for $\Gamma$ is cumbersome. For $\kappa_{L}=\kappa_{R}=\kappa$, and $\omega_{L}=\Omega$, it reads as
\begin{widetext}
\begin{equation}
\Gamma = \frac{2\kappa\Omega^{2}c^4(e^{x}-e^{y})}{-c^2(1+e^{-x}-2e^{x}-e^{2x}+3e^{-x+y})-2s^2(-1+e^{-x}+e^{x}-e^{-2x}-e^{y}+e^{-x+y})},
\end{equation}
\end{widetext}
where $x=\Omega/T_{L}$ and $y=\Omega/T_{R}$. The corresponding heat current expressions  for the unfiltered bath spectra are not presented here due to their cumbersome form.

There can be situations where BSF is not mandatory for rectification, as in Fig.~\ref{fig:AppC}, where the R bath is only coupled to the transition at the frequency $\Omega$ which is much less than the frequencies $\omega_L$, $\omega_L-\Omega$ and $\omega_L+\Omega$, whereas the L bath is coupled to the transitions with frequencies $\omega_L$ and $\omega_L+\Omega$. Therefore, when we change the bath temperatures, the cold bath temperature (here R bath temperature) is not sufficient to induce transitions
except at the frequency $\Omega$. 
Without BSF, the other possible global cycle is associated with the transitions at the frequencies $\omega_L$, $\omega_L-\Omega$ by the L bath and $\Omega$ by the R bath. For the parameter choice in Fig.~\ref{fig:qRect}, as the R bath temperature cannot induce the transition  even at the frequency $\omega_L-\Omega$ (as it is much higher than $\Omega$), there is no significant effect of BSF on the rectification. By contrast, for $\omega_L\sim\Omega$, the cold bath may able to induce  transitions at $\omega_L-\Omega$ that results in the decrease of heat rectification. In this case, BSF  drastically improves the rectification if we select our L bath spectrum such that, the transition $\omega_L-\Omega$ is either completely filtered out or at least drastically suppressed, as in the example of Fig.~\ref{fig:qqRectMechanism}. 

\section{\label{AppendixD} Amplification with BSF}
\label{ampli}
For the choice of bath spectra shown in Fig.~\ref{fig:qqAmpMech}(c), the master equation for the coupled TLS in the  diagonalized basis $\ket{j}$ is given by (in the notation of App. \ref{AppendixC})
\begin{eqnarray}
\mathcal{L}_{E}(\rho) &=& s^2\Gamma^{E}_{23}(\ket{3}\bra{3}-\ket{2}\bra{2}),  \nonumber\\
\mathcal{L}_{C}(\rho) &=& c^2\Gamma^{C}_{13}(\ket{3}\bra{3}-\ket{1}\bra{1})+c^2\Gamma^{C}_{24}(\ket{4}\bra{4}-\ket{2}\bra{2}), \nonumber\\
\mathcal{L}_{B}(\rho) &=& s^2\Gamma^{B}_{12}(\ket{2}\bra{2}-\ket{1}\bra{1})+s^2\Gamma^{B}_{34}(\ket{4}\bra{4}-\ket{3}\bra{3}).\nonumber\\
\end{eqnarray}
The steady-state heat currents are given by
\begin{eqnarray}\label{eq:AppHeatAmp1}
\mathcal{J}_{E} = \omega_{-}s^2\Gamma_{\text{T}}, \quad
\mathcal{J}_{B} = \Omega s^2\Gamma_{\text{T}},\quad
\mathcal{J}_{C} = -\omega_{a}s^2\Gamma_{\text{T}},\nonumber\\
\end{eqnarray}
where, $\omega_{-}=\omega_{a}-\Omega$, and note that the emitter and base heat currents are in the opposite direction to the collector current. The exact expression for $\Gamma_{\text{T}}$ is cumbersome. In the limit $T_{E}\to\infty$, it is simplified to the form
\begin{widetext}
\begin{eqnarray}\label{eq:appheat}
\Gamma_{\text{T}} &=&\frac{c^2 \gamma_{E}\gamma_{C}\gamma_{B}(e^{\Omega/T_{B}}-e^{\omega_{a}/T_{C}})}{c^2 \gamma_{E}\gamma_{B}(1+e^{\Omega /T_{B}})(1+e^{\omega_{a}/T_{C}}) + s^2\gamma_{E}[e^{\omega_{a}/T_{C}}\gamma_{B} +e^{\Omega /T_{B}}(\gamma_{E}+e^{\omega_{a}/T_{C}}(\gamma_{E}+\gamma_{B}))]},
\end{eqnarray}
\end{widetext}
where,
\begin{eqnarray}
\gamma_{E} = \omega_{-}\kappa_{E}\bar{n}(\omega_{-}), \quad
\gamma_{C} = \omega_{a}\kappa_{c}\bar{n}(\omega_{a}),\quad
\gamma_{B} = \Omega\kappa_{c}\bar{n}(\Omega). \nonumber\\
\end{eqnarray} 
\begin{figure}[t]
\centering
\hspace*{-0.60cm}
\includegraphics[width=0.5\textwidth]{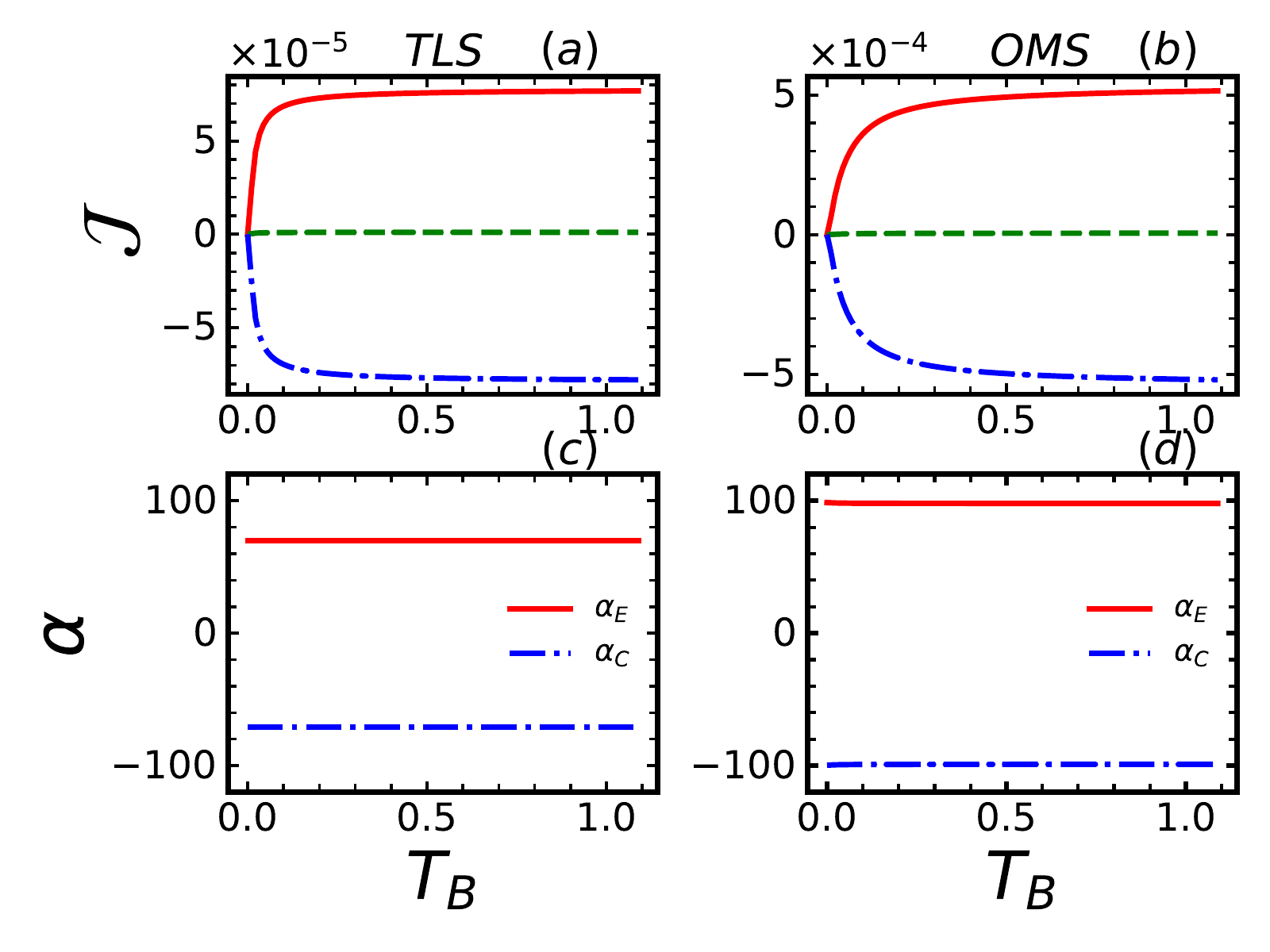}
\caption{(Color online)~{\it{Amplifier}}: (a), and (b) panels present the steady-state heat currents $\mathcal{J}$, (c) and (d)~amplification factors $\alpha$, as a function of the base temperature $T_{B}$, for the coupled TLS, and the OMS, respectively. In panels (a), and (b), solid, dashed, and dot-dashed lines are for $\mathcal{J}_{E}$, $\mathcal{J}_{C}$, and $\mathcal{J}_{B}$, respectively. In (c), and (d)
solid and dot-dashed lines are for $\alpha_{E}$ and $\alpha_{C}$, respectively, in addition, the 
emitter and collector baths have filtered spectral densities as shown in Fig.~\ref{fig:qqAmpMech}(c), and the base has FSD.  
 Parameters: $\omega_{a}=1$, $\omega_{b}=0.01$, $g=0.005$, $\kappa_{C}=\kappa_{E}=\kappa_{B}=0.001$, $T_{E}=1.1$, and $T_{C}=0.1$.}
\label{fig:AppD2}
\end{figure}
The amplification factors in this case evaluate to
\begin{eqnarray}\label{eq:AppAmpFact1}
\alpha_{E} = \frac{\partial\mathcal{J}_{E}}{\partial\mathcal{J}_{B}} =\frac{\omega_{-}}{\Omega}, \quad
\alpha_{C} = \frac{\partial\mathcal{J}_{C}}{\partial\mathcal{J}_{B}} =-\frac{\omega_{a}}{\Omega}.\quad
\end{eqnarray} 
For the selection of the baths spectra shown in Fig.~\ref{fig:AppD}(c), the possible transitions induced by the baths are shown in Fig.~\ref{fig:AppD}(a), and the steady-state heat currents in this case are given by
\begin{eqnarray}\label{eq:AppheatAmp2}
\mathcal{J}_{B}&=&\Omega s^2\Gamma_{1},\nonumber\\
\mathcal{J}_{E}&=&\omega_{-}s^2\Gamma_{1}-\omega_{a}c^2\Gamma_{2},\nonumber\\
\mathcal{J}_{C}&=& -\omega_{a}s^2\Gamma_{1} + \omega_{a}c^2\Gamma_{2},
\end{eqnarray}
where, $\Gamma_{1}:=\Gamma^{E}_{32}$ and $\Gamma_{2}:=\Gamma^{E}_{13}+\Gamma^{E}_{24}$. For the same system parameters, this choice of baths spectra allows us, as compared to  Fig.~\ref{fig:qqAmpMech}(c), more energy cycles to transfer heat between the baths, including direct transfer of heat between emitter and collector, which is indicated by slashed lines in Fig.~\ref{fig:AppD}(a). This can also be seen by comparing Eqs.~(\ref{eq:AppHeatAmp1}) and (\ref{eq:AppheatAmp2}), and noting that $\Gamma_{1}$ and $\Gamma_{2}$ have always opposite signs. To elaborate on this point, we consider an energy cycle shown in Fig.~\ref{fig:AppD}(b), which shows that $\Gamma^{C}_{24}>0$, and to transfer the heat directly between collector and emitter $\Gamma^{E}_{24}<0$. A similar cycle can be considered to show that if $\Gamma^{C}_{13}>0$ then $\Gamma^{E}_{13}<0$. Consequently, in all possible cycles, $\Gamma_{1}$ and $\Gamma_{2}$ must have opposite signs, which results in the increase of heat currents in the system compared to Eq.~(\ref{eq:AppHeatAmp1}). The amplification factors in this case is given by
\begin{eqnarray}\label{eq:AppAmpFact2}
\alpha_{E} = \frac{\omega_{-}}{\Omega}-\frac{\omega_{a}c^2}{\Omega s^2}\frac{\partial\Gamma_{2}}{\partial\Gamma_{1}}, \quad
\alpha_{C} = -\frac{\omega_{a}}{\Omega}+\frac{\omega_{a}c^2}{\Omega s^2}\frac{\partial\Gamma_{2}}{\partial\Gamma_{1}}.\quad
\end{eqnarray} 
Heat amplification is reduced for the choice of bath spectra shown in Fig.~\ref{fig:AppD}(c) as compared to the case presented in Fig.~\ref{fig:qqAmpMech}(c). This can be seen by comparing Eqs.~(\ref{eq:AppAmpFact1}) and (\ref{eq:AppAmpFact2}), and noting that $\partial\Gamma_{2}/\partial\Gamma_{1}>0$. In Fig.~\ref{fig:HeatAmplification}(a), we compare the amplification factors $\alpha_{E}$ for the baths spectra shown in Figs.~\ref{fig:qqAmpMech}(c),~\ref{fig:AppD}(c) and unfiltered baths, which shows that BSF can strongly increase the heat amplification. Similar results are obtained for the OMS.

\end{document}